\documentclass[12pt]{iopart}

\usepackage{cite}

\begin{document}

\title[Effective polar potential]{Comment on `Effective polar potential in
the central force Schr\"{o}dinger
equation'}
\author{Francisco M. Fern\'andez}

\address{INIFTA (UNLP, CCT La Plata-CONICET), Divisi\'on Qu\'imica Te\'orica,
Blvd. 113 S/N,  Sucursal 4, Casilla de Correo 16, 1900 La Plata,
Argentina}\ead{fernande@quimica.unlp.edu.ar}

\maketitle

\begin{abstract}
We analyze a recent pedagogical proposal for an alternative treatment of the
angular part of the Schr\"{o}dinger equation with a central potential. We
show that the authors' arguments are unclear, unconvincing and misleading.
\end{abstract}

In a recent paper Shikakhwa and Mustafa\cite{SM10} put forward an
alternative pedagogical discussion of the angular part of the solution to
the Schr\"{o}dinger equation for a quantum--mechanical model with a central
force:
\begin{equation}
-\frac{\hbar ^{2}}{2m}\nabla ^{2}\psi +V(r)\psi =E\psi
\label{eq:Schrodinger}
\end{equation}
They devoted part of the paper to show that this equation is separable in
spherical coordinates: $\psi (r,\theta ,\phi )=R(r)\Theta (\theta )e^{im\phi
}$, where $m=0,\pm 1,\ldots $, a discussion that appears in almost every introductory
textbook on quantum mechanics or quantum chemistry\cite{CDL77,EWK44}.

In particular, the authors concentrated on the polar equation
\begin{equation}
\frac{1}{\sin \theta }\frac{d}{d\theta }\sin \theta \frac{d\Theta }{d\theta }%
-\frac{m^{2}}{\sin ^{2}\theta }=-l(l+1)\Theta  \label{eq:polar}
\end{equation}
where $l=0,1,\ldots $ is the angular--momentum quantum number. They proposed
to convert this Sturm--Liouville equation into the Schr\"{o}dinger--like one
\begin{equation}
-\frac{1}{2}\frac{d^{2}y(\theta )}{d\theta ^{2}}+\frac{m^{2}-\frac{1}{4}}{%
2\sin ^{2}\theta }y(\theta )=Wy(\theta )  \label{eq:Shro_polar}
\end{equation}
where $y(\theta )=\sin ^{\frac{1}{2}}\theta \,\Theta (\theta )$ and $%
W=W_{l}=(1/2)[l(l+1)+1/4]$. We want to call the reader's
attention to the misleading notation used by the authors who called $E$ to
the eigenvalue of this equation as if it where the energy of the
central--field model (\ref{eq:Schrodinger}) (see their equations (12) and
(14)). To avoid such misunderstanding we choose the symbol $W$ for the
eigenvalue of the polar equation. We think that for a pedagogical discussion
it would have been more reasonable that the authors had chosen a rigid
rotator in which case the eigenvalue of the polar equation is proportional
to the energy of the system. It is worth adding that the transformation of a
Sturm--Liouville problem like (\ref{eq:polar}) into a Schr\"{o}dinger
equation like (\ref{eq:Shro_polar}) is well--known since long ago.

If we rewrite $|m|=0,1,\ldots ,l$ as $l=|m|+n$, $n=0,1,\ldots $ then we
derive the correct form of the eigenvalue of the polar equation in terms
of $m$ and $n$
\begin{equation}
W_{l}=\frac{1}{2}\left( l+\frac{1}{2}\right) ^{2}=\frac{1}{2}\left( n+|m|+%
\frac{1}{2}\right) ^{2}  \label{eq:W_l}
\end{equation}
which shows that $W_{l}$ does not depend on the sign of $m$, as expected
from the fact that the effective polar potential in Eq.~(\ref{eq:Shro_polar}%
) depends on $m^{2}$. On the other hand, the authors' polar energy $E_{n}^{m}
$ (see their equation (14)) depends on $m$ and does not clearly reflect the
degeneracy just mentioned. In order to derive their expression the authors
resorted to the following unconvincing and rather misleading argument
``These solutions are for non--negative $m$; those for negative $m$ are--as
is well known--directly proportional to these solutions.'' The
proportionality factor is in fact $e^{\pm i|m|\phi }$ and reflects part of
the degeneracy of the central--field models; for that reason one should not
neglect it so lightly.

The authors state that ``The solutions $\left| P_{l}^{m}(\cos \theta
)\right| ^{2}$ represent the probability of finding the particle at a
certain angle $\theta $''. They seemed to have forgotten the normalization
factor $N_{l}^{m}$ and that the polar part of the volume element is $\sin
\theta \,d\theta $ because the actual probability for finding the particle
between $\theta $ and $\theta +d\theta $ is well known to be $\left|
N_{l}^{m}P_{l}^{m}(\cos \theta )\right| ^{2}\sin \theta $. Besides, $\left|
P_{l}^{m}(\cos \theta )\right| ^{2}$ is not a \textit{solution} to the polar
equation.

The eigenvalue $W_{l}$ increases with $|m|$ as shown by Eq.~(\ref{eq:W_l}).
In order to explain this behaviour Shikakhwa and Mustafa\cite{SM10} plotted
the polar potential for increasing values of $|m|$ and showed that the
minimum increases. We agree that this graphical procedure is illustrative,
but the well--known Hellmann--Feynman theorem\cite{CDL77} is more elegant
and rigorous, and should be added to the discussion. If we consider the
eigenvalue equation $\hat{A}y=Wy$ ($y(0)=y(\pi )=0$) for the operator
\begin{equation}
\hat{A}=-\frac{1}{2}\frac{d^{2}}{d\theta ^{2}}+\frac{\lambda }{\sin
^{2}\theta }  \label{eq:A_op}
\end{equation}
where $\lambda $ is real, then that theorem states that
\begin{equation}
\frac{dW}{d\lambda }=\left\langle \sin ^{-2}\theta \right\rangle >0
\label{eq:HFT}
\end{equation}
Clearly, as $|m|$ increases $\lambda $ increases and $W_{l}$ increases.

Summarizing, we think that the paper by Shikakhwa and Mustafa\cite{SM10} is
not suitable for pedagogical purposes for the following reasons: first, they
apparently mistook the eigenvalue of the polar equation for the total energy
of the central field model, second, the treatment of the polar equation (\ref
{eq:Shro_polar}) is unclear and misleading. In particular, the polar
eigenvalue in their equation (14) does not clearly reveal the degeneracy
coming from the sign of $m$, and the argument for the restriction to $m\geq 0
$ in their expressions for the polar eigenvalue and eigenfunction is
unconvincing and unnecessary. In fact, the correct result follows
straightforwardly from the form of $W_{l}$ as we have already shown above.
Besides, the same simple argument clearly shows that $P_{l}^{|m|}(\cos
\theta )=P_{|m|+n}^{|m|}(\cos \theta )$ which is consistent with the
textbook treatment of the problem\cite{CDL77,EWK44}. We should also add the
sloppy discussion of the probability of finding the particle in a given
region of space.


\begin{thebibliography}{9}
\bibitem{SM10}  Shikakhwa M S and Mustafa M 2010 \textit{Eur. J. Phys.}
\textbf{31} 151.

\bibitem{CDL77}  Cohen-Tannoudji C, Diu B, and Lalo\"{e} F 1977 \textit{%
Quantum Mechanics} (John Wiley \& Sons, New York).

\bibitem{EWK44}  Eyring H, Walter J, and Kimball G E 1944 \textit{Quantum
Chemistry} (John Wiley \& Sons, New York).
\end{thebibliography}
\end{document}